# Terahertz probing of anisotropic conductivity and morphology of CuMnAs epitaxial thin films

P. Kubaščík[1], A. Farkaš[1,2], K. Olejník[2], T. Troha[3], M. Hývl[2], F. Krizek[2], D. C. Joshi[1], T. Ostatnický[1], J. Jechumtál[1], E. Schmoranzerová[1], R. P. Campion[4], J. Zázvorka[1], V. Novák[2], P. Kužel[3], T. Jungwirth[2,4], P. Němec[1], and L. Nádvorník[1]

1. Faculty of Mathematics and Physics, Charles University, Ke Karlovu 3, Prague 2 12116, Czech Republic
2. Institute of Physics of the Czech Academy of Sciences, Cukrovarnická 10, Prague 6 16200, Czech Republic
3. Institute of Physics of the Czech Academy of Sciences, Na Slovance 2, Prague 8 18200, Czech Republic
4. School of Physics and Astronomy, University of Nottingham, NG7 2RD, Nottingham, United Kingdom

## Abstract

Antiferromagnetic CuMnAs thin films have attracted attention since the discovery of the manipulation of their magnetic structure via electrical, optical, and terahertz pulses of electric fields, enabling convenient approaches to the switching between magnetoresistive states of the film for the information storage. However, the magnetic structure and, thus, the efficiency of the manipulation can be affected by the film morphology and growth defects. In this study, we investigate the properties of CuMnAs thin films by probing the defect-related uniaxial anisotropy of electric conductivity by contact-free terahertz transmission spectroscopy. We show that the terahertz measurements conveniently detect the conductivity anisotropy, that are consistent with conventional DC Hall-bar measurements. Moreover, the terahertz technique allows for considerably finer determination of anisotropy axes and it is less sensitive to the local film degradation. Thanks to the averaging over a large detection area, the THz probing also allows for an analysis of strongly non-uniform thin films. Using scanning near-field terahertz and electron microscopies, we relate the observed anisotropic conductivity of CuMnAs to the elongation and orientation of growth defects, which influence the local microscopic conductivity. We also demonstrate control over the morphology of defects by using vicinal substrates.

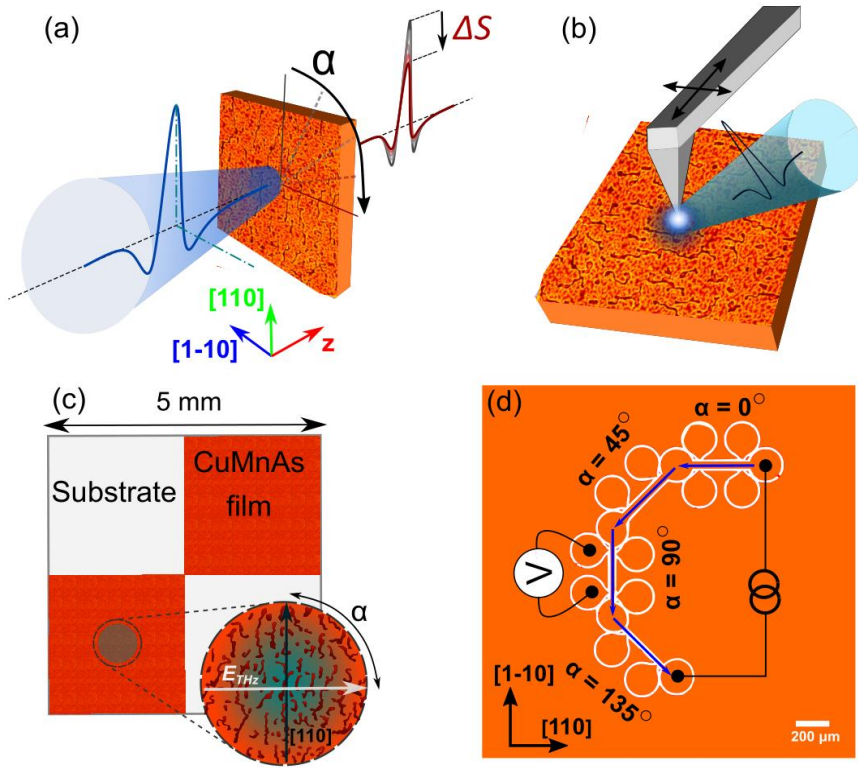

**Fig. 1 | Samples and experimental schemes.** (a) A schematic of the THz transmission probing of the conductivity $\sigma_{THz}$ of CuMnAs films. The incident linearly polarized THz pulse is transmitted through the film with [110] substrate crystal direction forming an angle $\alpha$ with the polarization plane. The transmitted beam has a reduced amplitude due to the local anisotropic $\sigma_{THz} = \sigma_{THz}(\alpha)$, resulting in a measured electrooptical signal $S(t)$. (b) A sketch of the SNOM setup where the THz pulse is focused on the sample surface and a scanning tip, resulting in a near-field interaction with resolution ~50 nm. The scattered THz radiation is related to the local $\sigma_{THz}$. (c) The chessboard pattern of CuMnAs films (orange areas) used for THz experiments. The drawing depicts the advantage of the contactless method: accessing the global characteristics by averaging local responses with an arbitrary angular resolution of $\alpha$ without an impact of inhomogeneity of the film. (d) A sketch of the lithographically prepared Hall bar device for DC electrical characterization. The electrical current (~100 µA) is sent through Hall bar segments (black arrows) oriented under four angles $\alpha = 0°, 45°, 90°$, and $135°$ with respect to [110] substrate crystal direction; the measured drop of potential provides the local resistance $R(\alpha)$.

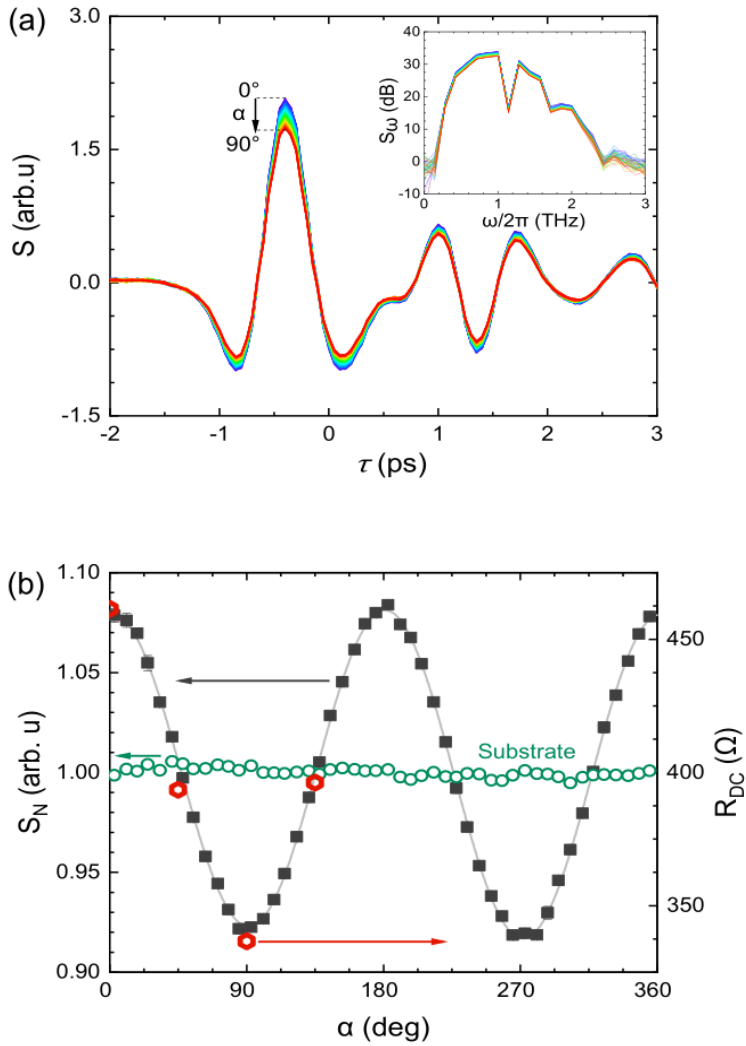

**Fig. 2 | THz transmission signals.** (a) Typical THz transmission traces $S(t)$ for various rotation angles $\alpha = 0\ldots 90°$ in 20-nm-thick CuMnAs on GaP. Inset: Corresponding spectra $S_\omega$. (b) Left axis: extracted RMS amplitudes $S_N(\alpha)$ of traces transmitted through the sample (full black squares) and bare substrate (green open circles) as a function of $\alpha$. Average of $S_N$ over $\alpha$ is normalized to unity for both data sets. The fit (grey curve) is $\Delta S\cos[2(\alpha + \alpha_0)] + 1$, yielding $2\Delta S = 16.6\%$ and $\alpha_0 = 1.5°$. Right axis: DC electrical resistance $R_{DC}$ (red hexagonal points) measured using the Hall bar device.

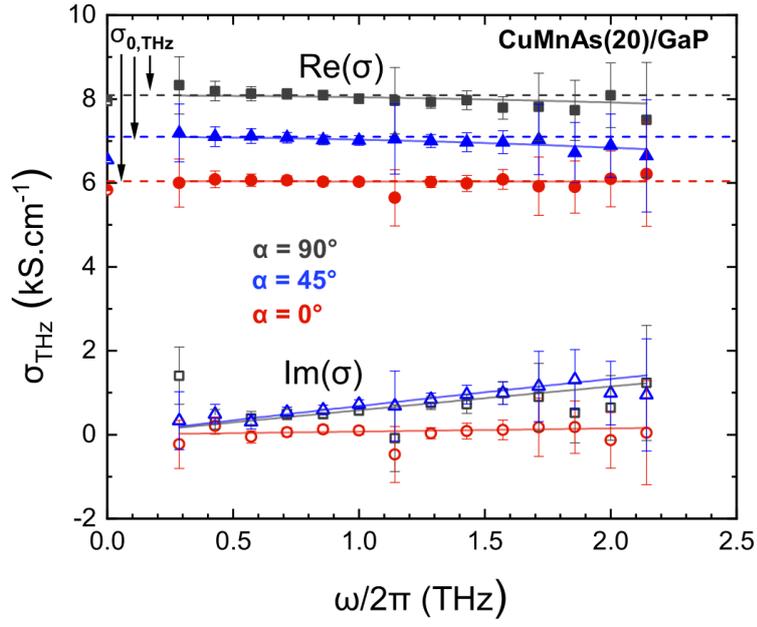

**Fig. 3 | THz conductivity.** Extracted real (full symbols) and imaginary parts (open symbols) of THz conductivity $\sigma_{THz}(\omega)$ for 20-nm-thick CuMnAs on GaP using Eq. (1) for three different orientations $\alpha$ of the sample. The solid curves are fits by the Drude model [Eq. (2)]. The value of the fit parameter $\sigma_{0,THz}$ is obtained from Eq. (2) and plotted by dashed horizontal lines. The data points at $\omega = 0$ are the DC conductivities $\sigma_{DC}$.

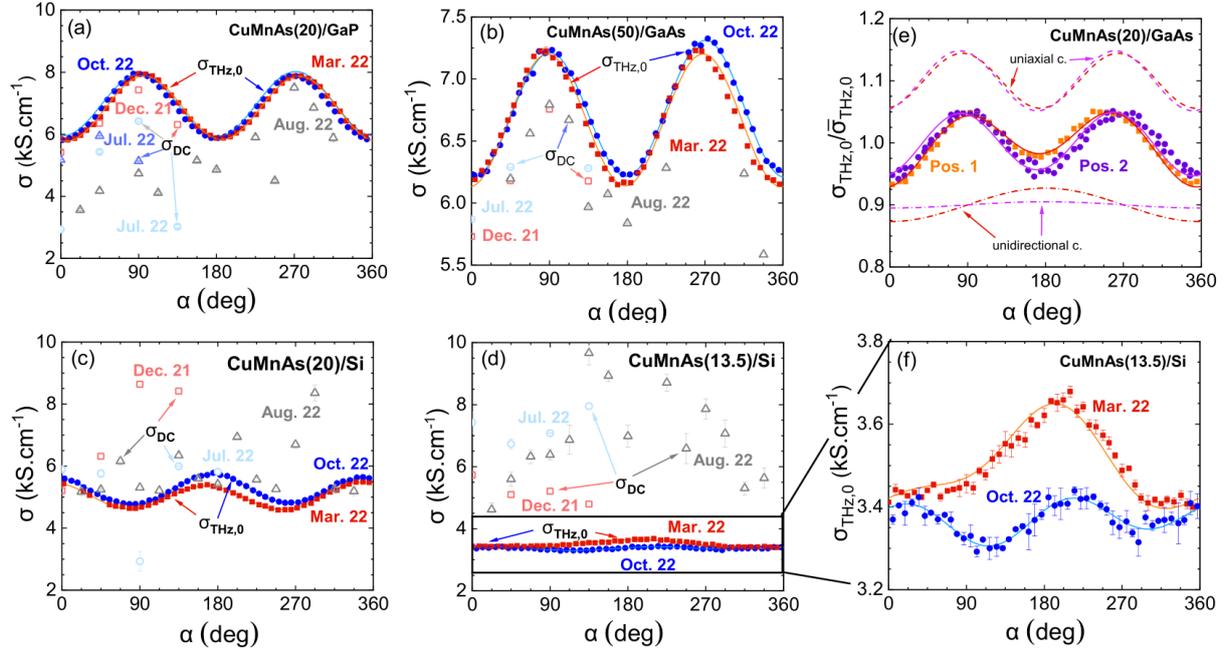

**Fig. 4 | THz vs. DC contrast of conductivity.** $\sigma(\alpha, \omega)$ obtained from the THz data ($\sigma_{\text{THz},0}$, full symbols) and DC measurements ($\sigma_{\text{DC}}$, open symbols) for various CuMnAs samples. (a) 20 nm thick CuMnAs on GaP substrate, (b) 50 nm on GaAs, (c) 20 nm on Si and (d) 13.5 nm on Si, measured at various times (see time labels in the plots). Solid curves are fits by Eq. (3). (e) $\sigma_{\text{THz},0}$ measured with 20-nm-thick CuMnAs on GaAs substrate on two different areas, separated by ~ 1 mm (in the horizontal [1-10] substrate direction for $\alpha = 0$). Dashed curves are the two-fold, uniaxial contributions of the fit by Eq. (3), the one-fold, unidirectional contributions are shown in dash-and-dot curves. These components are shifted vertically for clarity. (f) A close-up on $\sigma_{\text{THz},0}$ data of (d).

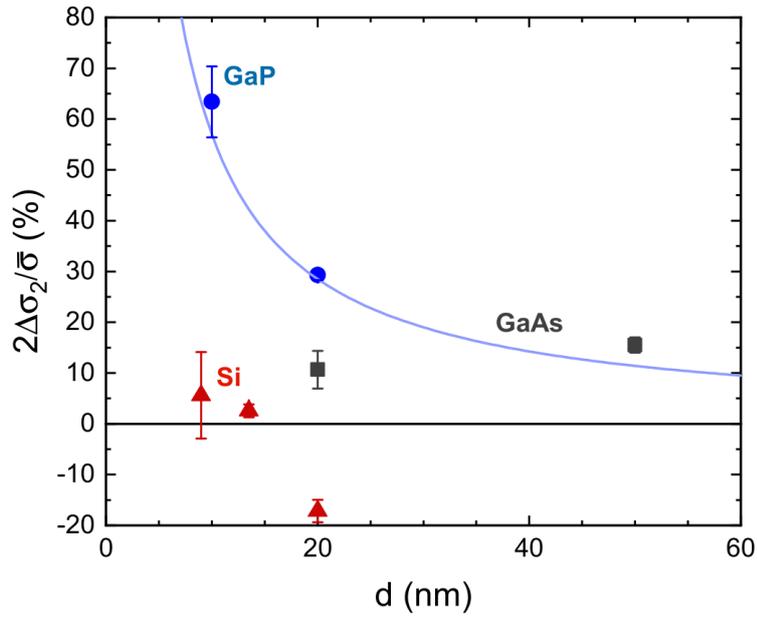

**Fig. 5 | Thickness dependence of conductivity anisotropy.** Conductivity contrast of the two-fold component $2\Delta\sigma_2/\bar{\sigma}$ as a function of the film thickness $d$ for three substrates (circles: GaP, squares: GaAs, triangles: Si) obtained from fits of THz data (Mar. 22) in Fig. 4. Error bars are the magnitude of the one-fold component $\pm\Delta\sigma_1/\bar{\sigma}$. In correspondence with Ref. [8], a $1/d$-dependence is plotted as guides for an eye for GaP (solid blue curve) and GaAs substrate (grey curve) and a constant function for Si substrate (solid red line). The grey dashed line stresses the rising trend with $d$ for GaAs substrate experimentally observed in our data. The opposite phase of $\sigma_{0,\text{THz}}(\alpha)$ for CuMnAs(20)/Si is stressed by plotting it as a negative number.

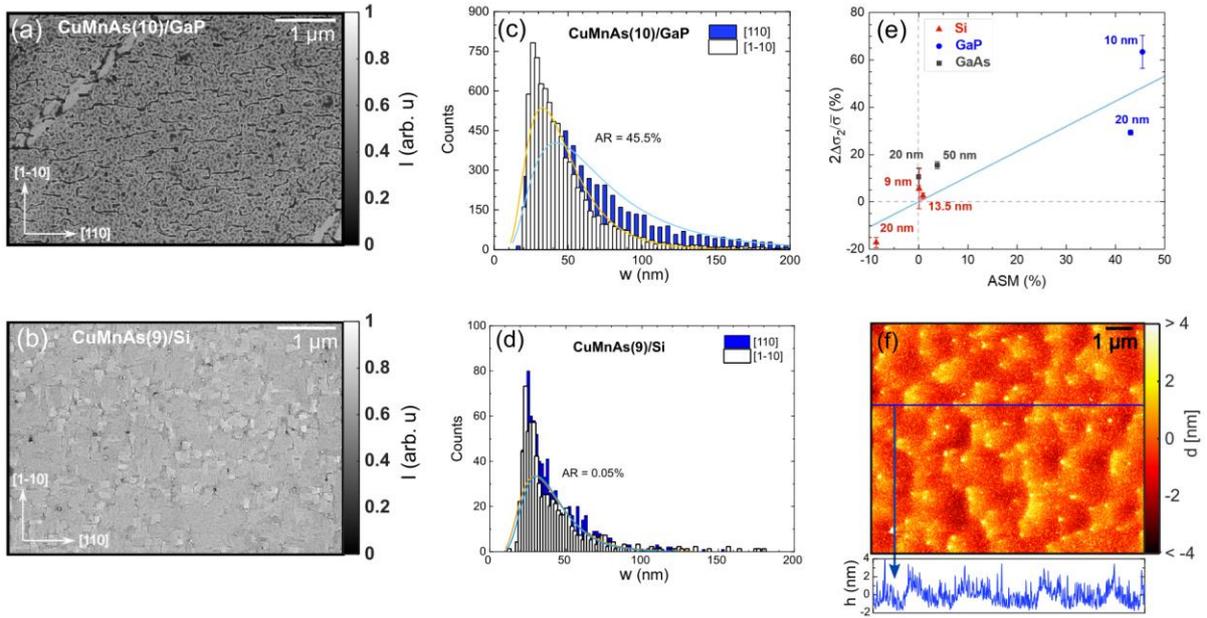

**Fig. 6 | Correlation of THz conductivity contrast with defect morphology.** SEM images of (a) CuMnAs(10)/GaP and (b) CuMnAs(9)/Si films with (c, d) corresponding histograms of defect size ($w$) distributions in [110] and [1-10] substrate directions. (e) Two-fold component of the conductivity contrast $2\Delta\sigma_2/\bar{\sigma}$ as a function of the anisotropy of the surface morphology $ASM$ of defects. The blue line is a guide for the eye. Error bars are considered as $\pm\Delta\sigma_1/\bar{\sigma}$. (f) An AFM image of the etched-off substrate part of CuMnAs(20)/Si sample, probing terrace-like surface structures.

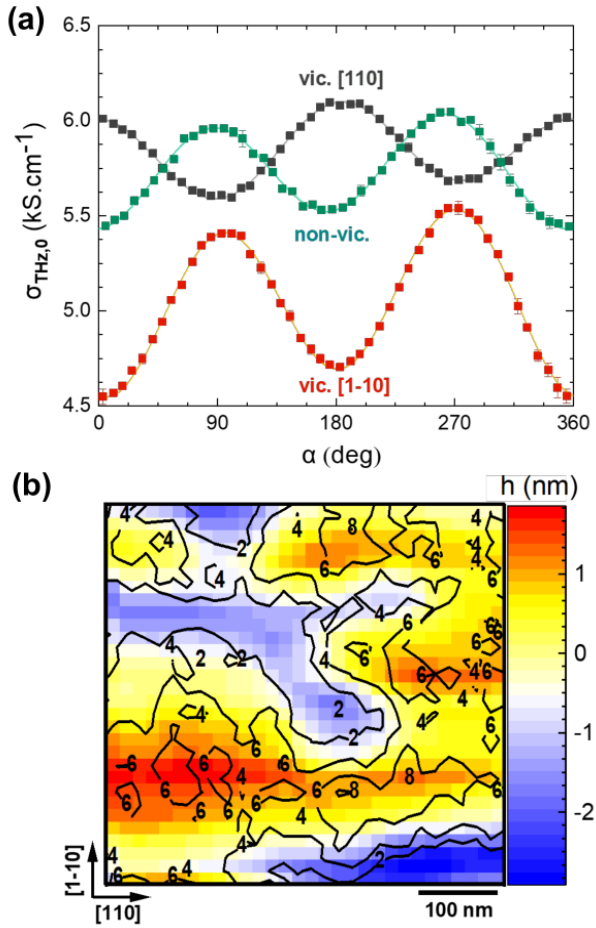

**Fig. 7 | Manipulation of conductivity contrast by vicinal substrates and THz-SNOM measurements**. (a) Anisotropy of $\sigma_{\mathrm{THz},0}$ in CuMnAs(20)/GaAs samples grown on normal (green points) and vicinal substrates with vicinality of 2° in direction [110] (black points) and [1-10] (red points). Curves with the corresponding colors are fits by Eq. 3. (b) The relative amplitude decrease in percent of the scattered THz pulses from CuMnAs(10)/GaP measured by the THz-SNOM scanning of the sample surface (contours with labels). The corresponding AFM image of the same area is shown by the color map.

# 1. Introduction

Antiferromagnets (AF) have appeared on the roadmap towards future spintronic applications in the information memory technology thanks to their advantages as compared to ferromagnets: their insensitivity to external magnetic fields up to units or tens of T improving the memory robustness[1,2], no stray fields, and the associated possibility of a denser memory bit integration[3], significantly faster coherent dynamics of the magnetic order reaching or exceeding THz frequency scale[4], and availability of a large variety of AF materials ranging from insulators to metals[5,6].

However, the lack of a net magnetic moment of antiferromagnets also brings a significant challenge for the concept of room-temperature device operation as impractically strong magnetic fields would be needed to manipulate the AF order. In this regard, the room-temperature collinear antiferromagnet CuMnAs has gained significant attention thanks to a theoretical prediction[7] and recent demonstrations of a coherent switching of the Néel vector by electrical current pulses[8–11]. Besides its coherent reorientation, the magnetic order of CuMnAs can be also manipulated incoherently through heat-activated processes induced by electrical[12], optical[2], and THz pulses[13,14], resulting in a ~~transient~~ metastable nano-fragmentation of AF domains[2,15,16]. The nano-fragmentation is accompanied by a switching of the resistive state by tens of percent, and has been explored in the context of logic-in-memory devices for spiking neuromorphic applications[2,12].

Consistently with previous studies[11-14], we prepare the tetragonal CuMnAs films (space group P4/nmm) by molecular beam epitaxy (MBE) on GaP, GaAs, and Si substrates[17,18]. A detailed characterisation[19] of MBE-grown CuMnAs showed that the crystal growth is accompanied by a formation of several types of defects associated with externally controlled parameters, such as stoichiometry, choice of substrates, and layer thickness $d$. For example, polar GaP and GaAs substrates promote a growth mode that starts from a formation of isolated elongated islands. Upon increasing the deposition time, the islands gradually evolve into a percolated layer containing an array of elongated holes whose density decreases with increasing $d$. In contrast, the shape of islands formed during the growth on the non-polar Si substrate remains isotropic for all $d$.

All the types of defects contribute to film inhomogeneities and induce local variations of conductivity. In addition, the elongated defects also create local morphologic anisotropy which can explain the observed global anisotropy of conductivity of CuMnAs thin films[14,19]. Moreover, it could also contribute to the magnetic anisotropy[20] and, thus, be linked to the efficiency of the switching of the magnetic order. Therefore, a convenient detection method, a detailed understanding, and even a manipulation of the morphologic anisotropy are desirable for improving the overall performance of CuMnAs-based devices and could bring new functionalities.

In this paper, we show that a reliable contactless non-destructive detection of the elongated-defect-related anisotropy of electrical conductivity of CuMnAs is possible by means of the time-domain THz spectroscopy. This method probes the electrical properties averaged over a large film area defined by the THz spot size (~1 mm²). These are orders of magnitude larger dimensions than the typical defect size (< 1 $\mu$m). Benefiting from this low sensitivity to local layer inhomogeneities and the high azimuthal angle resolution of the method, we find a correlation between the conductivity anisotropy and the anisotropy of elongated defects of defects. In addition, we show that Si substrates with vicinal (step-like) surfaces can be used to control the morphologic and conductivity anisotropy. Finally, by employing a scanning near-field THz microscopy, we directly observe local variations of the electrical conductivity in elongated defects, providing a microscopic insight into the origins of the global anisotropy of the conductivity in CuMnAs.

## 2. Experimental details

*2.1. Samples and setups*

The sample set consists of thin films of CuMnAs of various thicknesses (9-50 nm) grown by MBE on the three different (001)-oriented substrates (in most cases double-side polished): Si, GaP and GaAs, as summarized in the Supplementary material, Table S1. Since our aim is to study the defect morphology associated with elongated islands formed in the initial stages of the growth, we intentionally select samples in which this initial island growth-mode was well pronounced. The [100] crystal direction of the CuMnAs epilayer is oriented along the [110] direction of the substrate[19]. In the following text and Figures, the marked crystallographic directions refer to the substrate. The CuMnAs films were capped by 3 nm of Al which almost fully oxidized in the air and formed a protective $AlO_x$ cap. The surface was then lithographically patterned according to the experimental scheme used to determine the film conductivity (see Fig. 1 and Section 6: Experimental methods for more details).

The main, contactless technique used in our work is the THz time-domain transmission spectroscopy[21] [scheme shown in Fig. 1(a)], which is an established method for probing electrical currents[22–25] and conductivity[26,27]. Here, linearly polarized picosecond pulses of THz radiation, generated by an excitation of a spintronic emitter[28,29] by a train of ultrashort optical pulses (see Section 6 for more details), are focused to a ~860-$\mu$m-wide spot (see Supplementary Fig. S1). The radiation is further transmitted through the CuMnAs film and substrate, with the substrate crystallographic axis [110] rotated by an angle $\alpha$ with respect to the polarization direction of the THz pulse. The transmitted radiation $E(t)$ is detected by the electrooptical sampling resulting in a signal $S(t)$. Since its image in the frequency domain, $S(\omega)$, equals the product of the transmitted field $E(\omega)$ and the transfer function of the setup, we can relate the detected signals to the conductivity $\sigma(\omega)$ of the CuMnAs film by the Tinkham formula[21,22,30]:

$$\sigma(\omega, \alpha) = \frac{n_S(\omega) + n_A}{Z_0 d}\left(\frac{1}{t'(\omega, \alpha)} - 1\right), \tag{1}$$

where $n_S(\omega)$ and $n_A = 1$ are refractive indices of the substrate and air, respectively, $Z_0 \approx 377\ \Omega$ is the vacuum impedance, $d$ is the thickness of CuMnAs, and $t'(\omega, \alpha) = S(\omega, \alpha)/S_{\text{ref}}(\omega, \alpha)$. Here, $S_{\text{ref}}$ is the electrooptical signal for transmission through the bare substrate (see Section 6 and Supplementary Note 1 for more details). For switching between $S$ and $S_{\text{ref}}$, a chessboard-like pattern was created by non-etched and etched squares of the film [Fig. 1(c) and Experimental methods for details] to directly access the reference transmission through the substrate.

To get an additional insight into the local variations of conductivity, a THz scanning near-field optical microscopy (THz-SNOM) setup[31] was employed (see Sec. 6 for details about the setup). Although the scattered THz electric field depends on complicated near-field interactions between the sample and the SNOM tip, it can provide a qualitative relation between the growth defects, their topology, and the local conductivity of the CuMnAs film.

As a reference, the DC conductivity is determined by conventional electrical 4-probe measurements on lithographically defined Hall bars (length 250 μm, width 50 μm) patterned along selected crystallographic directions [Fig. 1(d)], including the [110] and [1-10] directions where the anisotropy of conductivity is the most pronounced. The Hall bars are distributed over an area comparable to the dimensions of the THz spot (~1 mm$^2$), however, each individual Hall bar probes only ~1% of this area.

Measurements by all techniques are performed at room temperature.

## 3. Results

*3.1 Raw THz signals*

Fig. 2(a) shows typical THz transmission waveforms $S(t)$ through a 20-nm-thick CuMnAs film on a GaP substrate, and the corresponding spectra. When we rotate the polarization direction from [110] axis of the substrate, i.e., increase α from 0 to 90°, the amplitude is clearly reduced. Since the overall shape of the THz trace is not changed, we can explore the α-dependence of the signal amplitude by taking the root-mean-square (RMS) of each waveform $S(t)$, normalizing its offset (average over α) to unity and plotting it in Fig. 2(b). The observed two-fold symmetry of the RMS amplitude $S_N(\alpha)$ is emphasized by fitting the data by $\Delta S \cos[2(\alpha + \alpha_0)] + 1$; the fit yields a modulation of $2\Delta S = 16.6\%$ and $\alpha_0 = 1.5°$.

To show that the amplitude variation originates uniquely in the CuMnAs layer, we compare the data to RMS of the transmission signals through the bare substrate [Fig. 2(b), green data, α-average rescaled to unity], which show a maximal variation of 0.6%. In addition, the symmetry of the signal is different, approaching a one-fold (unidirectional) pattern. Another verification of the origin of the modulation is provided by conventional DC electrical characterization of the thin film. The hexagonal symbols in Fig. 2(b), right axis, represent the resistance $R_{DC}$ measured in segments of the Hall bar oriented under an angle α. The two-fold modulation and the phase of the signals agree well with the THz data, and the value of $R_{DC}$ along the [110] direction is consistent with electrical measurements, reported previously on similar layers[19].

From the above qualitative findings, we conclude that the observed modulation of the THz signal originates in the thin film and is directly related to the anisotropic conductivity of CuMnAs. Moreover, the modulation has a phase consistent with the orientation of previously reported elongated surface defects (their elongated direction was observed to be parallel to the [110] crystallographic direction of the substrate) [19].

*3.2 THz and DC conductivities*

To address the quantitative comparison of both methods, we first extract the complex-valued conductivity, $\sigma_{THz}(\omega)$, from the THz data for $\alpha = 0, 45, 90°$ using Eq. (1) and correct it for possible phase shifts due to a variation of the substrate thickness [see also Supplementary Note 1]. We fit the data by the Drude model[32]:

$$\sigma_{THz}(\omega) = \frac{\sigma_{THz,0}}{1 - i\omega\tau}, \quad (2)$$

where $\sigma_{THz,0}$ is the conductivity at $\omega = 0$, and τ is the electron scattering time. The data and the fits are shown in Fig. 3. Apart from a significant variation of $\sigma_{THz,0}$ versus α (by ~30%) due to the anisotropy, we observe that the real part of $\sigma_{THz}$ remains nearly constant over the studied frequency range. This implies that (i) we can view the film as fully percolated with no significant portion of insulated islands of CuMnAs. In non-percolated films, $\sigma_{THz}(\omega)$ typically differs from the Drude behavior and would be manifested by an increase of $\text{Re}(\sigma_{THz})$ with increasing frequency[27,33]. Also, (ii) the real part of $\sigma_{THz}(\omega)$ can be approximated by constant $\sigma_{THz,0}$. The conclusions hold also for other samples (see Supplementary Fig. S2). We can directly compare $\sigma_{THz,0}$ to the DC conductivity $\sigma_{DC} = l/(wdR_{DC})$ where $l$ and $w$ are the length and width of Hall bar segments between the voltage contacts (data points at $\omega = 0$ in Fig. 3).

The conclusion (ii) allows us to avoid extraction of the spectral dependence of the conductivity and to simplify the subsequent systematic anisotropy study by disregarding the time dependence of $S(t)$ and only evaluating its RMS, $\text{rms}[S(t)]$. The quantity $\sigma_{THz,0}$ is then obtained by using Eq. (1) and substituting $t' = \text{rms}[S(t)]/\text{rms}[S_{\text{ref}}(t)]$. The full dependence of $\sigma_{\text{THz},0}$ and $\sigma_{DC}$ on $\alpha$, is plotted in Fig. 4(a) by full red and open red data points, respectively. We observe a good quantitative agreement between the two methods, yielding the same conductivity anisotropy $2\Delta\sigma/\bar{\sigma} = 29.3\%$, where $2\Delta\sigma$ is the full amplitude of modulation of conductivity and $\bar{\sigma}$ is the α-averaged mean conductivity.

An example of the analysis of samples on other substrates is shown in Fig. 4 [$S_N(\alpha)$ and $\sigma_{\text{THz},0}(\alpha)$ for the whole sample set are available in Supplementary Fig. S2 and S3, respectively]. In contrast to GaP and GaAs substrates [Fig. 4(a,b)], we observe a significant discrepancy between $\sigma_{\text{THz},0}$ (full red points) and $\sigma_{DC}$ (open red points) for CuMnAs films grown on the Si substrate [Fig. 4(c,d)]. To address the possible impact of an inhomogeneity and gradual oxidation of the films on both methods, we repeated the experiments after 7-8 months on the same samples: at a different location on the surface (THz data, full blue points), at the same location using the original Hall bar devices (DC data, blue open points) and on a new location using newly patterned devices (DC data, grey open points). We observe that, apart from CuMnAs(50)/GaAs, all DC measurements differ significantly from each other and even their scatter increased over the time period, while the THz data are significantly more consistent. These findings indicate that the DC electrical detection of the σ anisotropy is more prone to the local degradation of CuMnAs due to oxidation, an effect which was reported even on capped films[19]. The oxidation affects more likely the edges and holes of the film, leading to possible damages of the devices and deterioration of their functionality. The THz detection does not suffer from the oxidation to this extend as it averages over the area of the sample probed by the THz spot. For this reason, only the THz data will be analysed in the reminder of this section.

Although noticeably more reproducible, the THz data still show a certain deviation from the expected $\cos(2\alpha)$ dependence with indications of different symmetry with respect to α. Therefore, we evaluate the relative contribution of the uniaxial (two-fold, $\Delta\sigma_2$) and unidirectional (one-fold, $\Delta\sigma_1$) components by fitting the THz data by an empirical function

$$\sigma_{\text{THz},0}(\alpha) = \Delta\sigma_1 \cos(\alpha + \alpha_1) + \Delta\sigma_2 \cos[2(\alpha + \alpha_2)] + \bar{\sigma} \tag{3}$$

where $\alpha_1$ and $\alpha_2$ are the corresponding phases of modulations, i.e., orientations of the anisotropic axes for unidirectional and uniaxial components, respectively. Results of the fits of all the studied samples are summarized in Supplementary Table S1 and later in Fig. 5. While $\Delta\sigma_2 \gg \Delta\sigma_1$ and phase $\alpha_2 \approx 90°$ stays constant for all films on GaP and GaAs substrates, this does not hold for the Si substrate: the phase reverses its sign in Fig. 4(c), and the weaker anisotropy of $\sigma_{\text{THz},0}$ can even be dominated by $\Delta\sigma_1$, as seen in Fig. 4(d) and its close-up Fig. 4(f). Here, we see that $\Delta\sigma_1$ changed dramatically with changing the probing position and the measurement time period. To get more insight into the nature of $\Delta\sigma_1$, we repeated the experiment with CuMnAs(20)/GaAs on the same day and in two different sample locations separated by roughly 1 mm in the [1-10] direction [Fig. 4e]. While $\Delta\sigma_2$ stays the same for both locations, the $\Delta\sigma_1$ component significantly changes. A consistent observation is made for the corresponding phases: $\alpha_1$ varies randomly for all measured samples (illustrated in Supplementary Fig. S4). From these observations, we draw a conclusion that the unidirectional (one-fold) symmetry is an artifact. A slight misalignment of the sample rotation axis and the propagation axis of the THz pulses can lead to an effective motion of the THz spot over the surface. Indeed, the characterization of the maximal possible displacement of the THz spot on the sample surface during the sample rotation yielded ~ 40 μm (Supplementary Note 2 and Fig. S5). This, in combination with large-scale gradients of the film and the substrate, can lead to an apparent non-zero $\Delta\sigma_1$. Therefore, the genuine anisotropy of the electrical conductivity seems to be only the uniaxial component.

The elongated-defect-related uniaxial anisotropy component of the conductivity, $2\Delta\sigma_2/\bar{\sigma}$, for different $d$ and substrates is shown in Fig. 5, complemented by a guide to the eye (blue curve) to highlight the expected trends for the samples on the GaP substrate[19]. Conservatively, the error bars are set as the magnitude of the corresponding unidirectional anisotropy component $\Delta\sigma_1/\bar{\sigma}$. We observe that $2\Delta\sigma_2/\bar{\sigma}$ measured on CuMnAs/GaP decreases with the film thickness, while the sample on GaAs yields an unexpected increasing trend. Similarly, relatively small values of $2\Delta\sigma_2/\bar{\sigma}$, approaching zero within the error bars, on the two CuMnAs/Si samples for $d < 14$ nm are contrasted by a 20-nm-thick CuMnAs/Si sample with $2\Delta\sigma_2/\bar{\sigma} = 17.2\%$ and an opposite phase to the films on GaP and GaAs. To stress this phase change, the value of $2\Delta\sigma_2/\bar{\sigma}$ is plotted with a negative sign in Fig. 5. We note that all samples were remeasured on multiple surface regions, consistently yielding the presented results. To understand these trends, we explored further the morphology of the film defects.

*3.3 Surface defect morphology*

The morphology of surface defects can be observed directly by performing a scanning electron microscopy (SEM) imaging of the sample surfaces and inferring the corresponding anisotropy of the surface morphology $ASM$. We illustrate the analysis on two samples that differ significantly in their conductivity anisotropy: 10-nm-thick CuMnAs/GaP ($2\Delta\sigma_2/\bar{\sigma} = 63.4\%$) and 9-nm-thick CuMnAs/Si ($2\Delta\sigma_2/\bar{\sigma} < 3\%$), whose SEM images are shown in Figs. 6(a) and (b), respectively. The SEM images for all samples are shown in Supplementary Fig. S6. A clear elongation of defects is observed along the [110] GaP substrate direction (indicating $ASM > 0$), while rather isotropic surface defects are evidenced in the film on Si substrate ($ASM \approx 0$).

To quantify $ASM$ of the defects, we numerically processed the SEM images by the edge-finding method, described in Section 6, yielding histograms of defect dimensions, $w$, in [110] and [1-10] directions [Figs. 6 (c) and (d)]. By fitting the distributions by the inverse Gaussian function[34], we inferred $AMS = 2(\mu_{[110]} - \mu_{[1-10]})/(\mu_{[110]} + \mu_{[1-10]})$ where $\mu_i$ are the means of the distribution in the respective direction $i$. The analysis yields $AMS = 45.5\%$ and $0.05\%$ for the CuMnAs/GaP and the CuMnAs/Si sample, respectively. The results of the analysis for all samples are summarized in Fig. 6(e), where their $2\Delta\sigma_2/\bar{\sigma}$ are plotted as a function of the corresponding defect $ASM$. The observed trend directly reveals a correlation between these two quantities [shown in Fig. 6 (e) by a guide to the eye]. Remarkably, the scaling between the conductivity anisotropy and $AMS$ holds also for the measurements on GaAs and Si substrates where we found unexpected trends of the conductivity anisotropy vs. thickness $d$ (Fig. 5). The increased $2\Delta\sigma_2/\bar{\sigma}$ in thicker CuMnAs/GaAs is, indeed, related to more elongated defects in this sample, and the surprisingly large conductivity anisotropy in CuMnAs(20)/Si also scales well with the determined $ASM$. Moreover, the reversed phase $\alpha_2$ of $\sigma_{\text{THz},0}(\alpha)$ in this sample is consistent with the elongation of defects in the [1-10] substrate direction (opposite to CuMnAs on GaP and GaAs), yielding $ASM < 0$. This motivates us to verify whether such a large $ASM$ of CuMnAs films on non-polar Si substrates can be related to the substrate parameters.

*3.4 Effect of vicinal substrates*

Inducing anisotropic properties of thin films by growth on vicinal substrates is a known technique used in MBE deposition[35]. For example, a significant anisotropy in the conductivity of isotropic metals[36], semiconducting quantum wells[37], or in the morphology of microstructural defects in semiconducting thin films[38] has been observed on miscut (100) Si substrates. To address whether an unintentional miscut of the used substrates could explain the negative and large $ASM$ of the CuMnAs(20)/Si film, the atomic force microscopy (AFM) imaging of the etched-off substrate part of the sample was performed in Fig. 6(f). Here, we can indeed observe small terraces-like features in the form of asymmetric "saw-like" modulation of the surface, indicating a possible small vicinality in the order of 0.1°. On the other hand, an AFM imaging performed on the substrate part of CuMnAs(9)/Si, where small and positive $ASM$ and $2\Delta\sigma_2/\bar{\sigma}$ were observed, shows a considerably different surface morphology of etched substrates (Fig. S7).

To further test the hypothesis, the sample set was complemented by three 20-nm-thick CuMnAs films grown in a series under identical conditions on two intentionally miscut GaAs substrates of the vicinality of 2° in [110] or [1-10] directions and on a reference non-vicinal GaAs substrate. The obtained conductivity modulation $\sigma_{\text{THz},0}(\alpha)$ is plotted in Fig. 7(a). The sample grown on the reference non-vicinal GaAs substrate shows consistently a similar modulation, $2\Delta\sigma_2/\bar{\sigma} = 9.1\%$ with $\alpha_2 = 95°$, as compared to the CuMnAs(20)/GaAs sample shown in Fig. 5 and grown in a different MBE chamber. In contrast, the sample on the [1-10]-vicinal substrate shows a considerably higher anisotropy ($2\Delta\sigma_2/\bar{\sigma} = 16.1\%$, $\alpha_2 = 93°$), caused by a formation of defects elongated in the [110] direction. Interestingly, the [110]-miscut substrate even leads to an anisotropy in the perpendicular direction ($2\Delta\sigma_2/\bar{\sigma} = 7.1\%$, $\alpha_2 = 4°$), surpassing the original anisotropy related to the polarity of the non-vicinal GaAs. From these findings, we conclude that the vicinality of substrates can affect the nucleation of defects during growth and it can even be used to control the resulting conductivity anisotropy.

*3.5 Local conductivity and morphology*

Finally, we extend the phenomenological correlation of the conductivity anisotropy and *ASM* of defects to a more microscopic understanding. The local conductivity of the CuMnAs(10)/GaP film was investigated by THz-SNOM, yielding both the AFM-like maps of the surface, shown in Fig. 7(b) as color maps, and the amplitude of the scattered THz pulses, presented as contours with labels indicating the relative THz amplitude decrease in percent. We observe a clear correlation between the defects of the film and the locally scattered THz amplitude. If we limit ourselves only to a qualitative observation, we can interpret the findings as a clear correspondence between the surface morphology and the local conductivity, confirming the defect-related origin of the macroscopic anisotropy of the conductivity in CuMnAs films.

## 4. Discussion

The presented experiments show that the time-domain THz spectroscopy is a reliable, fast, and versatile method to quantify the anisotropic conductivity of CuMnAs and, potentially, of other thin films. The extracted angular modulation of the conductivity is consistent over multiple repeated measurements and with conventional DC electrical characterization. Compared to the DC probing using Hall bars, it brings several advantages: (i) It is non-destructive and contactless as it requires no surface patterning. (ii) It provides a high angular resolution of practically arbitrary precision, which is limited in the case of DC characterization by the number of patterned devices, allowing for a precise determination of anisotropic axes and exclusion of parasitic contributions of other than uniaxial (two-fold) symmetries in the signal. (iii) Since the THz probing of the conductivity is averaged over an arbitrarily large area according to a chosen spot size (in our case $\sim 1 \text{ mm}^2$), a single experiment can yield statistically averaged conductivity and its anisotropy even in strongly non-uniform samples.

The anisotropic conductivity of all the studied CuMnAs films, including samples on GaAs and Si substrates that did not follow the expected thickness dependence[19], was entirely explained by the induced anisotropy of the growth defects. Their anisotropy of the surface morphology, *ASM*, was directly determined from the SEM images and the local variation of the conductivity inferred from the THz-SNOM technique. Vice versa, the THz conductivity $\sigma_{\text{THz},0}(\alpha)$ was used for a fast and non-destructive determination of the orientation and *ASM* of the growth-induced defects. Indeed, the method revealed unexpected defect anisotropy on non-polar Si substrate, indicating an unintentional miscut in the used Si wafer.

The possibility of the control of the induced defect $ASM$ and related anisotropic conductivity using intentionally vicinal substrates can play an important role in the magnetic recording in CuMnAs. The shape and orientation of defects with a size on the order of tens of nanometers, i.e., the same range as observed in our samples [Fig. 6 (c)], were shown to affect also the magnetic anisotropy in AF. They lead to an effective shape-induced magnetic anisotropy[20], even though AF possess vanishingly small net magnetization. Indeed, very recently, Reimers *et al.* directly experimentally showed by photoemission electron microscopy and X-ray diffraction techniques that a magnetic domain structure in CuMnAs thin films is correlated with their growth defects[39].

## 5. Conclusion

In summary, we have shown that THz time-domain spectroscopy can be used to measure the growth-defect related anisotropic conductivity of CuMnAs thin films in a reliable, contactless, and versatile manner. The method also proved to be less sensitive to the film degradation as compared to the electrical characterization by Hall bars, requiring a lithographical processing. At the same time, it allows for an analysis of strongly non-uniform thin films due to the averaging of the conductivity over the large area of the THz spot size. The uniaxial conductivity anisotropy was observed on samples with various thicknesses and grown on GaP, GaAs, and Si substrates. Its magnitude and orientation with respect to the crystallographic directions in the sample were fully explained by the elongation and orientation of growth defects and related to the corresponding THz-SNOM detected local change of the conductivity. We demonstrated that the anisotropic conductivity can be controlled not only by the film thickness but also by using vicinal Si substrates. The understanding and control of defect morphology, presented in this paper, can lead to a future optimization of spintronic functionality in antiferromagnetic devices based on CuMnAs.

## 6. Experimental methods

*6.1. Hall bar fabrication*

For the fabrication of the chessboard like structure [Fig.1(c)] and the Hall-bar devices, first, a positive photoresist was deposited on the sample by spin-coating it at 4000 RPM and baked on the hotplate. Next, the resist was exposed using Microwriter ML3 PRO, developed using an alkaline developer, and etched in a 1:400 solution of phosphoric acid.

*6.2 Electrical measurements*

Electrical measurements were carried out with two electrical setups, one consisting of Keithley SourceMeter 2400 (current source) and Keithley MultiMeter 2000 (voltmeter) and the other utilizing NI DAQ card USB-6211. In the first setup, the readout current was set to 100 µA to prevent heating, and, in the case of the second setup, the reading voltage was set to 0.2 V, and the readout current was measured as a voltage drop over a known resistor. Corresponding voltages were measured with a four-point method for each segment in the Hall bar. The conductivity in the corresponding direction was calculated from the known dimensions of segments.

*6.3 THz measurements*

*Time domain THz spectroscopy*

The linearly polarized picosecond pulses of THz radiation were generated by exciting the spintronic emitter[29] by a train of laser pulses (1030 nm, 170 fs long, fluence 0.15 mJ/cm$^2$, repetition rate 10 kHz) and focused on the sample, forming a ~ 800-$\mu$m-wide spot (FWHM). The transmitted radiation $E(t)$ was detected by the electrooptical sampling using a 110-cut 2-mm-thick GaP crystal[21], yielding the electrical signal $S(t)$. Since its image in the frequency domain, $S(\omega)$, equals the product of the transfer function of the setup and the transmitted field $E(\omega)$, the transfer function cancels in ratio $S(\omega, \alpha)/S_{\text{ref}}(\omega, \alpha) = E(\omega, \alpha)/E_{\text{ref}}(\omega, \alpha) = t'$ where $S_{\text{ref}}$ and $E_{\text{ref}}$ are the electrooptical signal and electric field for transmission through the bare substrate.

To study the anisotropy of transmission, the sample was attached to a rotational holder with an angular resolution of 0.2°. If the THz beam axis and the axis of rotation of the holder do not coincide, small displacements of the beam spot on the sample surface are induced during the sample rotation. To avoid this, a special alignment technique was introduced (see Supplementary Note 2). After the alignment, samples were installed with a precision ~2° into the rotational holder without moving it. The measured angular dependence of THz transmission was globally shifted by 3° due to the THz polarisation rotation tilt from the vertical direction induced by parabolical mirrors (see Fig. S8).

*THz-sSNOM*

A THz scanning near-field optical microscope THz-SNOM (Neaspec) was used to investigate local changes in conductivity. Broadband THz pulses (0.2-1.5 THz) are focused on a PtIr tip (50 nm radius) which oscillates above the film surface with the typical tapping frequency of 50-100 kHz [Fig. 1(b)]. The detected scattered THz electric field carries information about the local conductivity. Scanning the sample surface with the tip thus provides local conductivity measurements with a spatial resolution of ~ 50 nm. The simplest models used to describe the underlying physics consider the electrostatic approach. In this case, the near-field contribution to the scattered THz electric field can be expressed as $E_{\text{scatt}} \propto (1 + r_{\text{p}})^2 \alpha_{\text{eff}} E_{\text{inc}}$, where $E_{\text{inc}}$ is the incident field, $\alpha_{\text{eff}}$ is the effective tip polarizability and $r_{\text{p}}$ the Fresnel reflection coefficient for *p*-polarized light.[40,41] The near-field interaction between the sample and the tip (and consequently the local conductivity) is incorporated in $\alpha_{\text{eff}}$ which depends nonlinearly on the tip-sample distance. As a result of the tip oscillation $\alpha_{\text{eff}}$ is modulated at the tip frequency as well as at higher harmonics of the tip frequency. By demodulating the THz signal at higher demodulation orders, we can efficiently extract the useful signal, which depends on the local conductivity. A 2D scan of the sample surface [Fig. 7(b)] was obtained by scanning the sample surface at the peak electric field of the scattered THz pulse detected at the second demodulation order. The quantitative determination of the local conductivity requires an appropriate model for the description of the whole system. However, on a qualitative level, we can relate THz signals to variations of local conductivity.

*6.4 Microscopic imaging*

*Scanning electron microscopy*

Defects in CuMnAs lead to a modification of morphology of the film surface; thus, we studied the surface of the samples by SEM as well as by AFM. To analyse the observed defects, SEM image processing was performed. The used approach is based on looking for certain characteristics of the defects along the two orthogonal directions (x and y-axis). For this, several cuts were applied along the x and y directions of the recorded image. In between those cuts, a peak finder method was applied and the found peaks were fitted by a set of Gaussian distributions. From the obtained data sets, histogram analysis was performed, for more details, see Supplementary Note 3.

*Atomic force microscopy*

For all AFM scans, Bruker Dimension ICON AFM was operated in semi-contact Peak Force QNM mode using Aspire conical force modulation (CFM) probes. The symmetrical shape and combination of small tip radius (guaranteed $< 10$ nm) and sharp tip cone angle (30°) of these probes ensure true and symmetrical representation of all sample features. For processing the data, Gwyddion software[42] was used with a focus on local surface roughness, disregarding the natural slope of the sample surface.

## Acknowledgement

The authors acknowledge funding by the Czech Science Foundation through projects GA CR (Grant No. 21–28876J and 19-28375X), the Grant Agency of the Charles University (grants No. 166123 and SVV–2022–260590), the Operational Program Research, Development, and Education financed by the European Structural and Investment Funds and the Czech Ministry of Education, Youth and Sports (Project No. SOLID21 CZ.02.1.01/0.0/0.0/16/019/0000760). The authors also acknowledge CzechNanoLab Research Infrastructure supported by MEYS CR (LM2023051). J.Z. acknowledges the support of Charles University grant PRIMUS/20/SCI/018.

# Terahertz probing of anisotropic conductivity and morphology of CuMnAs epitaxial thin films

P. Kubaščík[1], A. Farkaš[1,2], K. Olejník[2], T. Troha[3], M. Hývl[2], F. Krizek[2], D. C. Joshi[1], T. Ostatnický[1], J. Jechumtál[1], E. Schmoranzerová[1], R. P. Campion[4], J. Zázvorka[1], V. Novák[2], P. Kužel[3], T. Jungwirth[2,4], P. Němec[1], and L. Nádvorník[1]

1. Faculty of Mathematics and Physics, Charles University, Ke Karlovu 3, Prague 2 12116, Czech Republic
2. Institute of Physics of the Czech Academy of Sciences, Cukrovarnická 10, Prague 6 16200, Czech Republic
3. Institute of Physics of the Czech Academy of Sciences, Na Slovance 2, Prague 8 18200, Czech Republic
4. School of Physics and Astronomy, University of Nottingham, NG7 2RD, Nottingham, United Kingdom

## *Supplementary Notes*

## Supplementary Note 1: Conductivity extraction using Tinkham formula

**Refractive indices.** The values of refractive indices, used in Eq. (1) in the main text, were approximated as constants 3.34[1], 3.6[2] and 3.42[3,4] for GaP, GaAs and Si, respectively. In the case of GaP and GaAs, the error of the neglected dispersion in the range 0.2– 2.5 THz is smaller than 1% (see Supplementary Fig. S11), which is significantly below our experimental uncertainty of 6%. Also, literature values from different works can differ by ~ 1% (for example the refractive index of GaP is reported 3.34 in Wei *et al.*[5] and 3.36 in Tanabe *et al.*[1]). In case of Si, there is no frequency dependence reported in the relevant frequency range[3,4].

**Validity of the Tinkham approximation.** The Tinkham formula[6,7] is a thin-film approximation for highly conductive materials. It is derived from the general expression of the transmission coefficient $t'$ through a thin film under the normal incidence

$$t'(\omega) = \frac{E(\omega)}{E_{\text{ref}}(\omega)} = \frac{2n(n_s + 1)\exp\left(-\frac{i\omega(n-1)d}{c}\right)\exp\left(-\frac{i\omega(d_s - d_r)(n_s - 1)}{c}\right)}{(n+1)(n_s + n) - (n-1)(n - n_s)\exp\left(-\frac{2i\omega nd}{c}\right)}, \quad (S1)$$

where $n$ is the refractive index of a thin conductive layer, $n_s$ is the refractive index of substrate, $d$ is the thickness of the layer, $d_r$ is the thickness of a reference substrate and $d_s$ is thickness of the substrate underneath the thin layer. In the first order of expansion in $nd$ (the thin film approximation), it follows

$$t'(\omega) = \frac{n_s + 1}{n_s + 1 + a + Z_0 \sigma d} \exp(-i\omega\Delta d(n_s - 1)/c), \quad (S2)$$

where $a = \omega n_s d/c$, $Z_0$ is the vacuum wave impedance, $\sigma$ is the conductivity of the thin layer and $\Delta d = d_s - d_r$. This approximation requires two conditions to be fulfilled: (i) $n_s \ll n$ and (ii) $a \ll 1$. To verify the first condition, we assume that there are no interband transitions in CuMnAs in our THz spectral range. Under this assumption, we estimate the refractive index of CuMnAs $n = \sqrt{1 - i\sigma/(\omega\varepsilon_0)}$, where $\sigma \approx 5.10^5$ S.m$^{-1}$, which yields $n = 150 - 150i$ to $n = 45 - 45i$ in spectral range from 0.2 to 2.5 THz, values significantly larger than refractive indices of Si, GaAs or GaP. To verify condition (ii), we calculated the value of the factor $a$ for the maximal film thickness $d = 50$ nm, yielding $a < 0.002$ at 2.5 THz which fulfils the condition. The error coming from the used approximation can be estimated as $\sigma_a = a/(Z_0 \sigma d) < 0.02\%$.

**Phase shift in raw data.** An additional phase shift in $t'(\omega)$ is expected due to inhomogeneity of the substrate thickness between the positions where $E$ and $E_{\text{ref}}$ are measured. This is described by an unknown exponential factor $\exp(-i\omega\Delta d(n_s - 1)/c)$ in Eq. (S2). The factor is determined by assuming that the conductivity follows the Drude model [Eq. (2)] and by minimizing the sum of squares during the fitting process by Eq. (2). Parameter $\Delta d$ together with other parameters of extracted THz conductivity are summarized in Table S2. To cross-check our assumption of the Drude behaviour, we also performed a complementary analysis using the Smith-Drude model

$$\sigma_{\text{THz}}(\omega) = \frac{\sigma_{0,THz}}{1 - i\omega\tau_{DS}}\left(1 + \frac{c_1}{1 - i\omega\tau_{DS}}\right), \quad (S3)$$

yielding $c_1$ of the order of $10^{-5}$ which confirms the validity of the Drude model. In the case of the sample CuMnAs(13.5)/Si, $c_1$ reaches -0.05 and -0.016 for $\alpha = 0$ and 90°, respectively

## Supplementary Note 2: Alignment of rotation and optical axes

**Determination of THz spot size.** The size of the THz spot size was determined by the scanning edge method: we measure the maximum of the THz waveform $S$ while shadowing the beam by a razor in the direction $x$ perpendicular to the propagation vector and we fit it by the analytical expression

$$S \propto \int_{-\infty}^{+\infty} \int_{-\infty}^{x} e^{-\frac{x^2+y^2}{w^2}} \mathrm{d}y \mathrm{d}x \propto 1 + \mathrm{erf}(x/w) \tag{S4}$$

where $w$ is 1/e-value of gaussian beam. After multiplication by $2\sqrt{\ln 2}$, it yields the full width at half maximum of the THz electric field.

**Alignment of axes.** To align the rotational axis of the holder and the optical axis of THz beam, we attached a razor to the holder and set it so that it cuts a half of the transmitted signal. The overlap of the axes was found by minimizing the variation of the transmitted signal upon a 180° rotation of the razor. The optimal position resulted in the maximal signal modulation of ~ 8.5 %, corresponding to ~ 40 μm misalignment. The displacement of the THz beam over the surface of samples during their rotation was, thus, smaller than 40 μm [see Figs. S1 and S5(a)]. To prove that no modulation is caused by the holder itself, we also show a measurement with an empty holder [Fig. S5(b)].

## Supplementary Note 3: Algorithm of histogram analysis and dependence on external parameters

The analysis of the SEM images is based on analysing of cuts in the vertical and horizontal directions. Such intensity profiles allow us to determine the width of defects $w$ in both relevant crystallographic directions ([110] and [1-10]) and reveal the anisotropy of their morphology at a chosen position. To suppress effect of noise, intensity profiles were smoothened by the Savitzky-Golay filter of rank 5 with the frame length 11. Consequently, a peakfinder method was applied and the found peaks corresponding to elongated defects were fitted by the Gaussian function.

The set of obtained $w$ (defined as 1/e-value of the Gaussian function) is consequently converted into a histogram. The obtained histogram distribution typically shows a significant skewness and, therefore, we fit it by the inverse Gaussian distribution. The result of the histogram analysis depends on the prominence – a relative height of the defects with respect to surrounding peaks. In the presented results [Fig. 6(c,d)], the prominence was kept constant at the value of 110.

## Supplementary Tables

**Table S1: Table of coefficients of the angular dependence of the conductivity according to Eq. (3) for the whole data set.** Phases $\alpha_1$ and $\alpha_2$ are set to lie in the interval between (-180°…180°) and (0°…180°), respectively.

| Substrate | $d$ (nm) | $\bar{\sigma}$ (kS.cm$^{-1}$) | $\sigma_2/\bar{\sigma}$ (%) | $\alpha_2$ (°) | $\sigma_1/\bar{\sigma}$ (%) | $\alpha_1$ (°) |
|---|---|---|---|---|---|---|
| Si | 9 | 2.94 | 5.6 | 27 | 8.5 | -82 |
|  | 13.5 | 3.37 | 2.7 | 41 | 2.1 | -126 |
|  | 20 | 5.24 | 17.2 | 175 | 2.2 | -102 |
| GaAs | 20 | 4.08 | 10.7 | 93 | 3.8 | -117 |
|  | 50 | 6.74 | 15.5 | 91 | 1.5 | -94 |
| GaAs vic. [110] | 20 | 5.74 | 7.1 | 4 | 2.2 | 137 |
| GaAs vic. [1-10] | 20 | 5.05 | 16.1 | 93 | 4.0 | 143 |
| GaAs non-vic. | 20 | 5.64 | 9.1 | 95 | 2.1 | 147 |
| GaP | 10 | 4.30 | 63.4 | 90 | 7 | -54 |
|  | 20 | 6.96 | 29.3 | 91 | 1.2 | 71 |

**Table S2: Summary of regression coefficients from the spectral dependence of the conductivity according to Eq. (2).** The last subscript $i$ in $\sigma_{\text{THz},0,i}$ corresponds to the angle of the sample rotation. The samples from the study of the impact of the substrate vicinality are marked by an asterisk.

| Substrate | $d$ (nm) | $\Delta d_0$ (μm) | $\Delta d_{90}$ (μm) | $\Delta d_{45}$ (μm) | $\tau_0$ (fs) | $\tau_{90}$ (fs) | $\tau_{45}$ (fs) | $\sigma_{\text{THz},0,0}$ (kS·cm$^{-1}$) | $\sigma_{\text{THz},0,90}$ (kS·cm$^{-1}$) | $\sigma_{\text{THz},0,45}$ (kS·cm$^{-1}$) |
|---|---|---|---|---|---|---|---|---|---|---|
| Si | 9 | 1.8 | 2 | 2.2 | 0.32 | 0.32 | 0.32 | 3.22 | 3.24 | 3.22 |
|  | 13.5 | -0.57 | -0.49 | -0.48 | 0.32 | 0.32 | 0.32 | 3.32 | 3.37 | 3.38 |
|  | 20 | 0.61 | 0.67 | 0.79 | 0.32 | 0.32 | 0.32 | 5.1 | 5.36 | 5.41 |
| GaAs | 20 | -0.59 | -0.38 | -0.46 | 1.4 | 0.32 | 1.8 | 4.47 | 3.2 | 4.36 |
|  | 50 | 0.092 | -1.1 | -0.72 | 0.32 | 19 | 13 | 7.23 | 6.28 | 6.8 |
|  | 20* non. vic. | 0.21 | -0.4 | -0.5 | 0.32 | 14 | 16 | 6.04 | 5.72 | 5.7 |
|  | 20* vic. [1 1 0] | 0.27 | 0.25 | 0.26 | 0.32 | 0.32 | 0.32 | 5.58 | 5.02 | 4.96 |
|  | 20* vic. [1 -1 0] | -1.3 | -1.4 | -1.3 | 0.32 | 0.65 | 0.67 | 5.71 | 5.87 | 5.86 |
| GaP | 10 | 0.37 | 0.89 | 0.64 | 0.32 | 0.32 | 0.32 | 5.74 | 3.03 | 4.41 |
|  | 20 | -1.5 | -0.11 | -0.67 | 25 | 11 | 18 | 8.12 | 6 | 7.25 |

*Supplementary Figures*

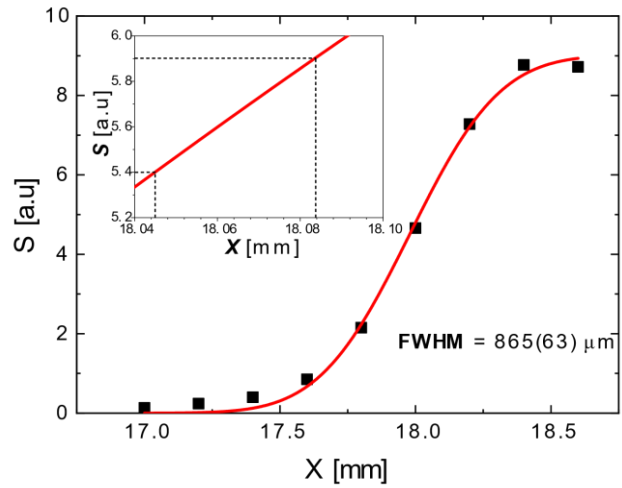

**Fig. S1: Scanning edge measurement and determination of beam size**. The EOS signal $S$ of the maximum of the transmitted THz waveform (points) as a function of the knife edge position $x$ with a corresponding fit by Eq. S4. The fit yields the full width at half maximum (FWHM) of $(865 \pm 63)$ µm. The inset shows a zoomed part for a variation of $S$ corresponding to the modulation observed with a rotating razor (cf. Fig. S4). The dashed lines show the maximal modulation of 8.5%, which yields an estimated mismatch of rotation axes ~40 µm.

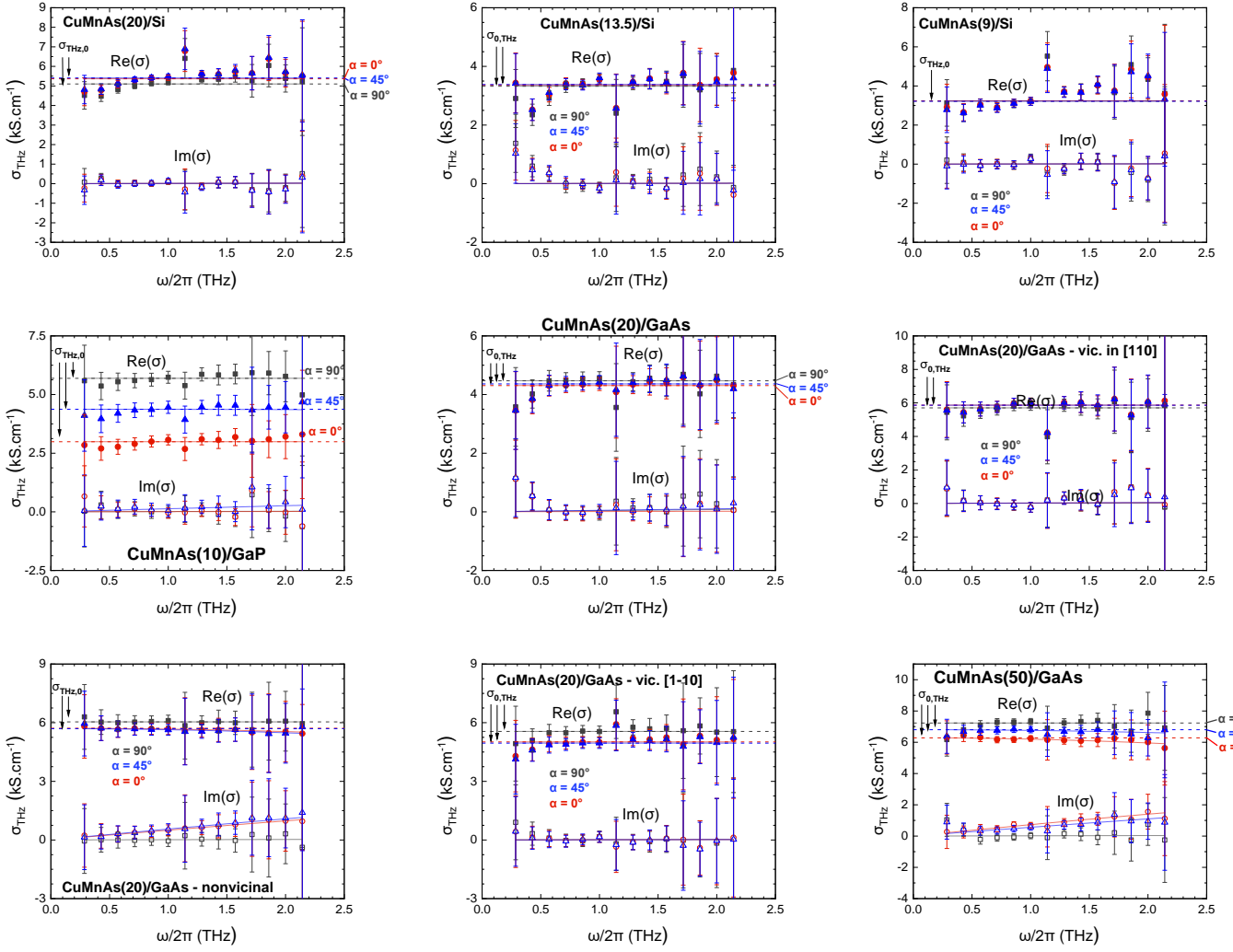

**Fig. S2:** THz conductivities $\sigma_{\text{THz}}$ extracted for the rest of the listed samples. Full and empty points correspond to the real and imaginary parts of $\sigma_{\text{THz}}$, respectively, and full lines are fits by the Drude model [Eq. (2)]. Dashed lines, corresponding to $\sigma_{THz,0}$, show that approximation by a constant function is legitimate. Regression coefficients corresponding to the presented figures are summarized in Table S2.

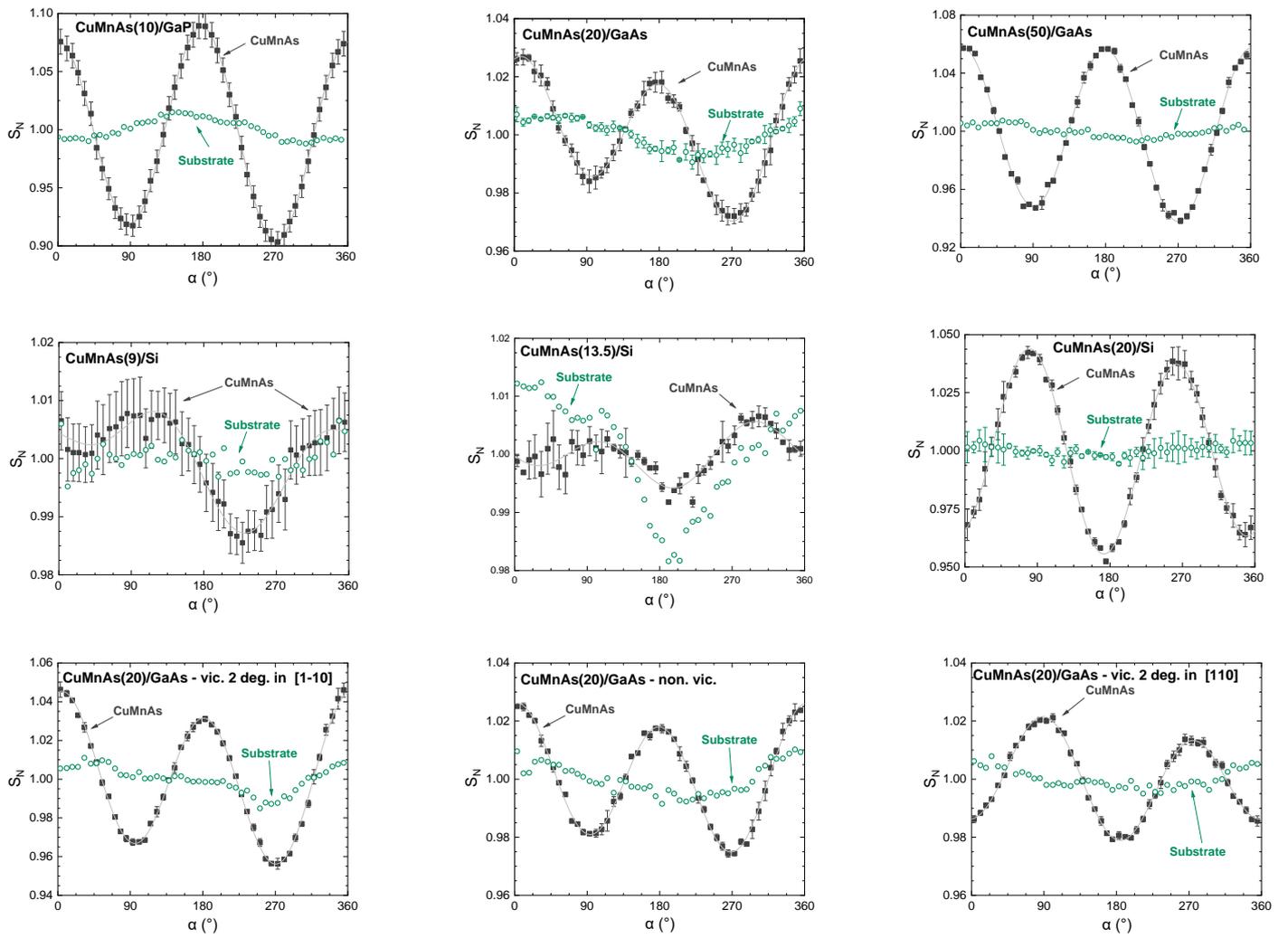

**Fig. S3: Modulation of THz signals for all the listed CuMnAs thin films**. Normalized THz transmission signal $S_N$ plotted as a function of angle α for all CuMnAs samples grown on different substrates. The unidirectional and uniaxial components were extracted by fitting the modulations to Eq. (3) and summarized in Table S1.

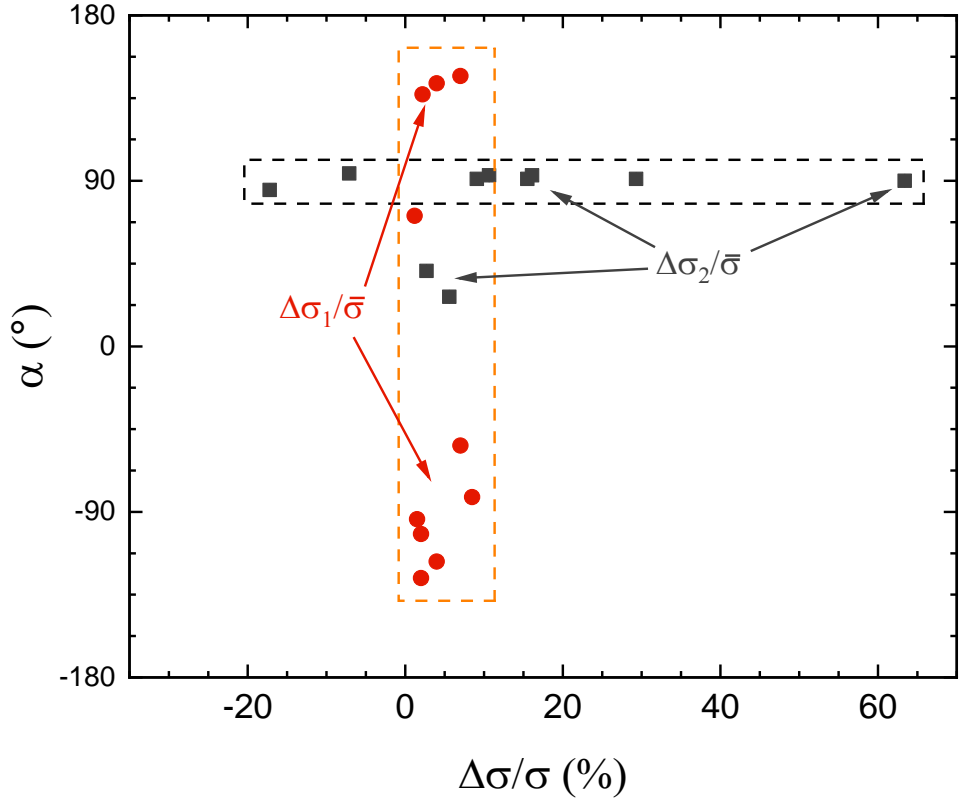

**Fig. S4: Amplitude and phase of modulation contrasts.** Grey and red points correspond to phases $\alpha_2$ and $\alpha_1$ plotted as a function of contrasts $\Delta\sigma_2/\bar{\sigma}$ and $\Delta\sigma_1/\bar{\sigma}$, respectively. $\alpha_2$ and $\alpha_1$ are set to lie in the intervals (0°…180°) and (-180° … 180°), respectively. In the case of sample CuMnAs(20)/Si, the sign was reversed and $\alpha_2$ shifted by 90 degrees to emphasize that defects are rotated by 90°. Notice that while $\Delta\sigma_2/\bar{\sigma}$ is characterized by a strong modulation and constant phase $\alpha_2$ (corresponding to [110] direction), phase $\alpha_1$ of $\Delta\sigma_1/\bar{\sigma}$ is significantly more random and modulation $\Delta\sigma_1/\bar{\sigma}$ is bellow 5%.

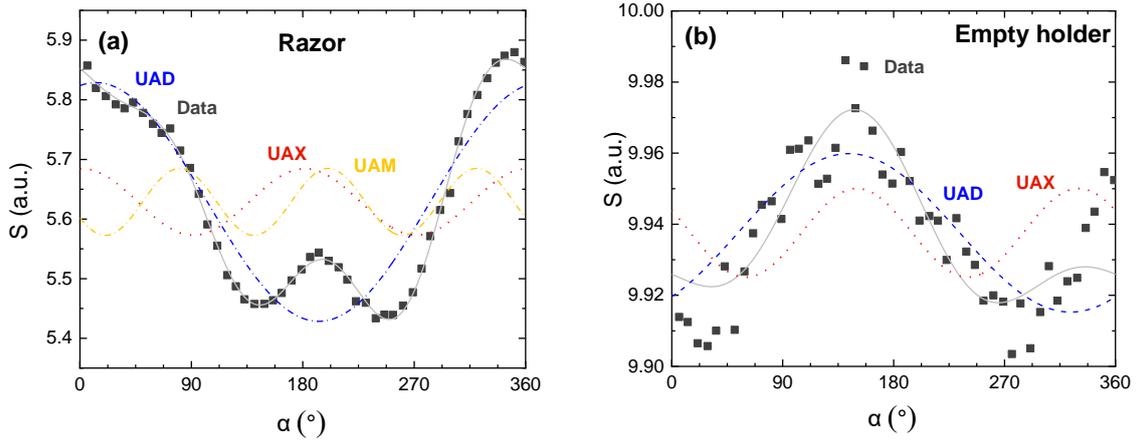

**Fig. S5: Overlap of optical axis of THz beam and rotation axis of holder.** a) The RMS of THz transmission signal $S(t)$ as a function of rotation angle α in the rotating razor method, detailed in Supplementary Note 2. The maximum observed variation of $S(t)$ is 8.5 %, whose major part (~7.2%) originates from the unidirectional component (UAD, $\propto \sin(\alpha)$). Apart from the UAD component, we also observe small contributions ($< 2\%$) corresponding to the uniaxial (UAX, $\propto \sin(2\alpha)$) and the mixed unidirectional-uniaxial components (UAM, $\propto \sin(3\alpha)$). The UAX and UAM components originate likely from the asymmetry of the THz beam spot. Using the measurement in Fig. S1(a) and its inset, we determine the maximal apparent movement of the beam over the sample surface of 40 μm. This value is considerably smaller than FWHM of the THz spot size (828 μm). The same measurement but with an empty holder is shown in b), demonstrating a negligible contribution of the rotating holder itself.

**Fig. S6**: SEM images of the CuMnAs sample set.

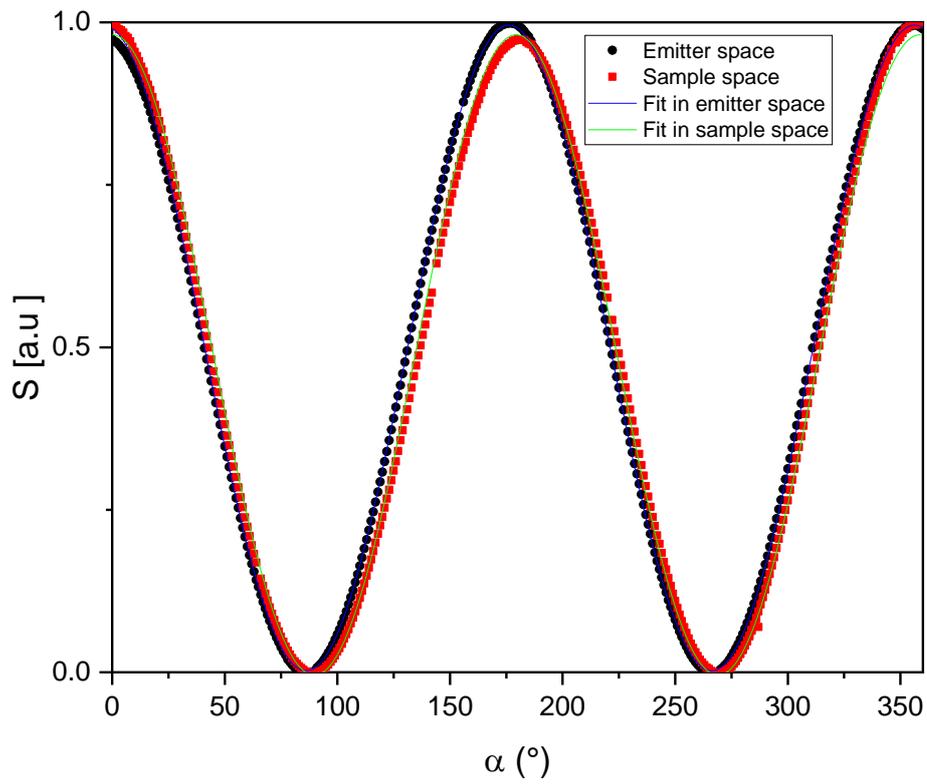

**Fig. S7:** Transmitted EOS signal $S$ through the wire grid polariser as a function of the polariser easy-axis rotation in the sample plane (black circles) and the emitter position (red squares). To estimate the polarisation tilt, results were fitted by $S(\alpha) = S_0 + \Delta S \cos[2(\alpha - \alpha_0)]$, leading to an approximately 3° tilt between the sample position (green line) and the emitter position (blue line).

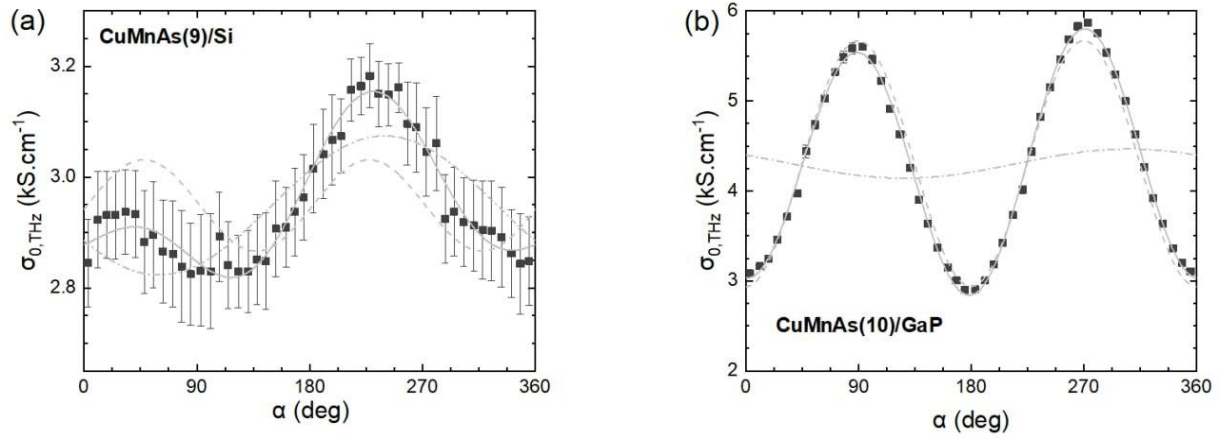

**Fig. S8: Angular dependence of THz conductivity for the rest of dataset.** Measured conductivity (dark grey points) for (a) CuMnAs(9)/Si and (b) CuMnAs(10)/GaP samples were fit by Eq. 3 (solid line) and anisotropy components $\frac{2\Delta\sigma_2}{\bar{\sigma}}$ and $\frac{2\Delta\sigma_1}{\bar{\sigma}}$ are marked by dashed and dash-dot lines, respectively.

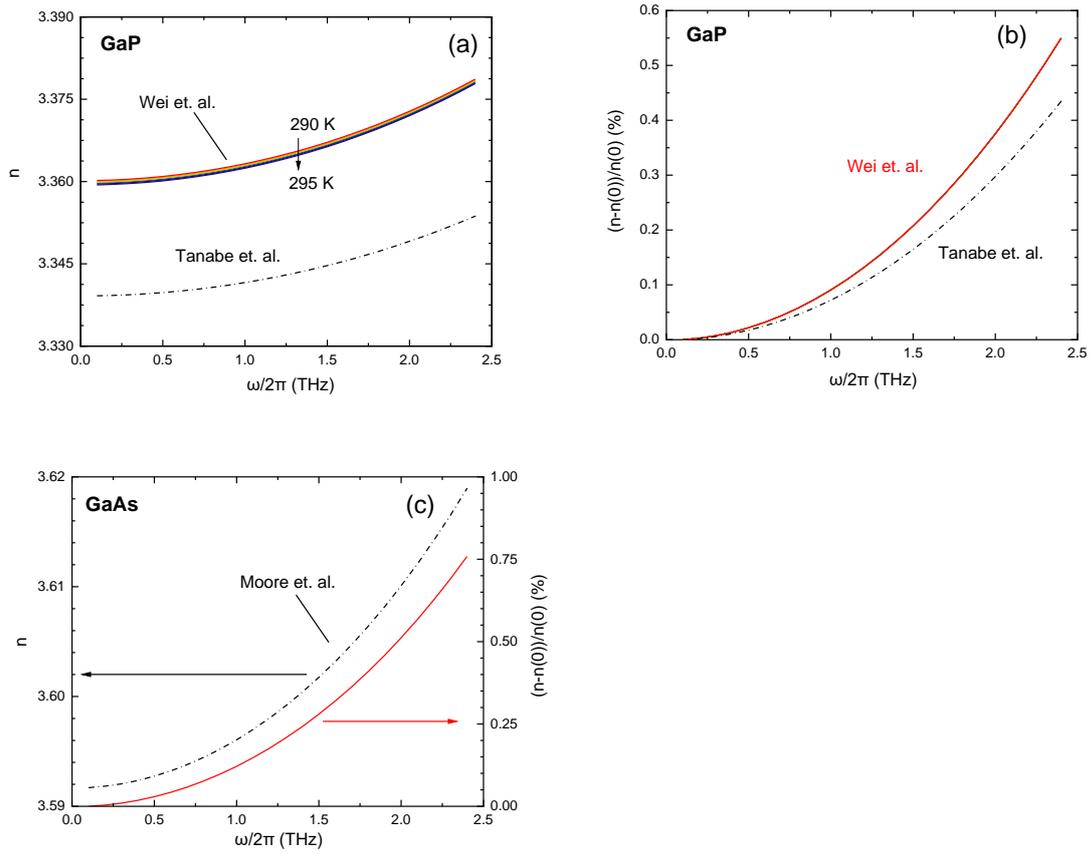

**Fig. S9: Dispersion of refractive indices of GaP and GaAs from literature.** (a) The dispersion relations of GaP in the THz spectral range according to Wei *et al*[5]. in the temperature range from 290 to 295 K (coloured lines) and Tanabe *et. al*[1]. (dashed black line). (b) The relative change in the refractive indices according to both references. (c) The dispersion relation of GaAs in the THz spectral range according to Moore et al[2].